\documentclass[prl,twocolumn,superscriptaddress,showpacs,amsmath,amssymb]{revtex4-1}

\usepackage{graphicx}
\usepackage{latexsym}
\usepackage{amsmath}
\usepackage{amssymb}
\usepackage{amsfonts}
\usepackage{color}
\usepackage{bm}% bold math
\usepackage{verbatim}

\begin{document}

 \title{Odd triplet superconductivity
 induced by the moving condensate }

 \date{\today}

\author{M.A.~Silaev}
\affiliation{Department of
Physics and Nanoscience Center, University of Jyv\"askyl\"a, P.O.
Box 35 (YFL), FI-40014 University of Jyv\"askyl\"a, Finland}
\affiliation{Moscow Institute of Physics and Technology, Dolgoprudny, 141700 Russia}

\author{I. V. Bobkova}
\affiliation{Institute of Solid State Physics, Chernogolovka, Moscow
  reg., 142432 Russia}
\affiliation{Moscow Institute of Physics and Technology, Dolgoprudny, 141700 Russia}
\affiliation{Dubna State University, Dubna, 141980, Russia}

\author{A. M. Bobkov}
\affiliation{Institute of Solid State Physics, Chernogolovka, Moscow reg., 142432 Russia}

%%% abstract
 \begin{abstract}
 It has been commonly accepted that  magnetic field suppresses superconductivity
 by inducing the ordered motion of Cooper pairs.  We demonstrate  that  magnetic field can instead  provide a generation of  superconducting correlations by inducing  the motion of superconducting condensate.
 This effect arises in superconductor/ferromagnet heterostructures in the presence of Rashba spin-orbital coupling.
  We predict the odd-frequency spin-triplet superconducting correlations called the Berezinskii order to be
  switched on at large distances from the superconductor/ferromagnet interface by the application of a magnetic field.
    This is shown to
 result in the unusual behaviour of Josephson effect and  local density of states in superconductor/ferromagnet structures.
 \end{abstract}

%%% PACS numbers
\pacs{} \maketitle

 The phenomenon of superconductivity dwells on the condensation of Cooper pairs each consisting of two electrons with opposite momenta.
 Magnetic field  tends to break Cooper pairs by
 inducing their center-of mass motion which makes the momenta of two paired electrons to be not exactly opposite. This seems to be the fundamental mechanism called the orbital depairing effect which exists in any superconducting system and  leads to the suppression of superconductivity by the magnetic field \cite{TinkhamBook}.

 %%%%%%%%%%%%%%%%%%%%%%%%%%%%%%%%%%%%%%%%%%%%%%%%%%%%
 \begin{figure}[h!]
 \centerline{$
 \begin{array}{c}
 \includegraphics[width=0.9\linewidth]{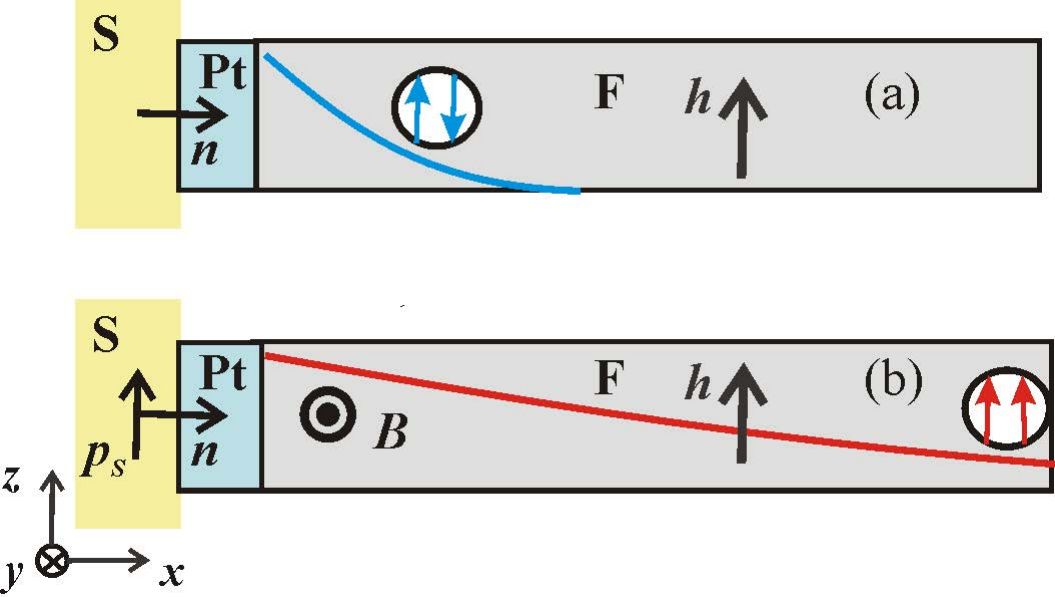}
 \end{array}$}
 \caption{\label{Fig:1}
   Schematic picture of the system considered:  diffusive superconductor/ferromagnet junction with Rashba SOC at the S/F interface induced by the thin heavy metal Pt layer. (a) only short-range superconducting correlations are present  in the absence of magnetic field. (b) Magnetic field generates condensate momentum through the Meissner effect
   $ \bm p_s   = - (\lambda_L/ \Phi_0)\bm n\times \bm B $. The interplay of condensate momentum $\bm p_s$, SOC and exchange field $\bm h$  leads to the generation of long-range s-wave spin-triplet component for $\bm B \neq 0$. }
 \end{figure}
 %%%%%%%%%%%%%%%%%%%%%%%%%%%%%%%%%%%%%%%%%%%%%%%%%%%%

 In this Rapid Communication we show that the magnetic field can in fact stimulate
 %superconducting properties such as the
 Josephson current by generating correlations which are odd in the frequency domain.
 Our proposal is based on the observation that the combination of
  exchange field, spin-orbit coupling (SOC) and
 controllable condensate motion is generically enough for the effective
 manipulation of the odd-frequency spin-triplet pairing states \cite{berezinskii1974new} which have attracted continual interest for several decades
 \cite{
belitz1992even,
belitz1999properties,
balatsky1992new,
abrahams1995properties,
coleman1994odd,
BergeretPRL2003,
bergeret2005odd,
fominov2015odd,
linder2019odd,
black2012odd,
black2013odd,
DiBernardo2015,
komendova2017odd,
cayao2017odd,
cayao2020odd,
triola2018odd,
cayao2018odd,
dutta2019signature,
sukhachov2019odd,
alidoust2017pure,
tanaka2007odd,
asano2007odd,
asano2013majorana,
yokoyama2008theory,
banerjee2018controlling}.
We find that the condensate motion
 generated by the external magnetic field through the Meissner effect
adds an additional degree of freedom which can control the generation of odd-frequency superconductivity in currently available experimental setups with SOC \cite{Satchell2018,Satchell2019,jeon2019tunable,
jeon2019abrikosov,
PhysRevB.99.024507,
PhysRevApplied.11.014061,
Jeon2018}.
In the context of superconductor (S)/ferromagnet (F) hybrid structures
\cite{oh1997superconductive,tagirov1999low,kadigrobov2001quantum,gu2002magnetization,fominov2003triplet,fominov2010superconducting,halterman2007odd,halterman2008induced,zhu2010angular,alidoust2014meissner}
these correlations are known as
long-range triplets (LRT) because they can penetrate at large distances into the ferromagnetic material \cite{Bergeret2001a,
Bergeret2001,
PhysRevB.76.060504,
bergeret2005odd,
keizer2006spin,
fominov2007josephson,
Braude2007,
eschrig2008triplet,
eschrig2003theory,
Robinson2010a,
Robinson2010,Singh2016,Khaire2010,Mironov2015,halterman2016half,alidoust2018half,PhysRevApplied.8.044008}.
We suggest a mechanism which leads to the
 stimulation of  long-range Josephson current in SFS junctions by the external magnetic field
 thus  opening great perspectives for low-dissipative spintronics \cite{Linder2015,Eschrig2015a}.

   Recently the phenomena employing SOC in the
SF structures has attracted much attention \cite{niu2012spin,
chung2011topological,
dimitrova2007theory,
mal2010inverse,
bobkova2017quasiclassical,
tkachov2017magnetoelectric,
mironov2017spontaneous,
alidoust2015long,
alidoust2015spontaneous,
hikino2018magnetization,
PhysRevB.92.125443,
Bergeret2013,
Bergeret2014,
Tokatly2017,
Bobkova2004,
Bergeret2015,
Campagnano2015,
Pershoguba2015,
Malshukov2016,
Malshukov2018,
Buzdin2008,
Konschelle2009,
robinson2019chirality,
Jeon2018,
jeon2019tunable,
Montiel2018}.
The SOC is generated at interfaces due to the broken inversion symmetry and usually has Rashba-type form
\cite{Rashba1959Symmetry, bihlmayer2015focus, ast2007giant,bihlmayer2006rashba,manchon2015new}
$(\alpha/m)( \hat{\bm \sigma}\times \bm p)\cdot \bm n$ where $\bm n$ is the interface normal,
$\hat{\bm \sigma}$ is the vector of spin Pauli matrices, $\bm p$ and $m$ are the electron momentum and mass.
It can be additionally enhanced by adding
a thin layer of heavy metal \cite{bihlmayer2006rashba,manchon2015new} as
 sketched in Fig.\ref{Fig:1}.
Usual values of interfacial Rashba level splitting in metals  can be assumed to be  small  $v_F \alpha /E_F \ll 1$, where $v_F$ and $E_F$ are the  Fermi velocity and energy, respectively.
Under this condition
the modification of spin-singlet pairing by the SOC alone is negligible \cite{Gor'kov2001,PhysRevB.67.020505,PhysRevB.92.125443,Reeg2015,bobkova2017quasiclassical}.  However, even in this case the combined effect of SOC and spin-splitting field shows up in the
quite efficient generation of spin-triplet superconducting correlations.

The combination of interfacial SOC and exchange field can induce p-wave spin-triplet superconducting correlations \cite{chung2011topological} which mediate the long-range Josephson effect through clean ferromagnets\cite{niu2012spin,PhysRevB.94.014515}.
%The p-wave correlations can be observed in  ballistic SF systems \cite{chung2011topological,niu2012spin,PhysRevB.94.014515}.
 %
In a diffusive system  p-wave correlations decay at the mean free path which is much smaller than the thickness of F layer.
At the same time p-wave correlations
 exist near the interface  with non-zero SOC. They can be converted  to  s-wave correlations due to the anisotropic Doppler shift of quasiparticle energy levels \cite{kohen2006probing} $\bm v_F\cdot\bm p_s$ induced by the condensate motion along the interface as shown in Fig.\ref{Fig:1}b.
This shows up in the generation of LRT pairs which are robust to both  disorder and exchange interaction.
  This effect can be engineered  combining usual superconductors such as Al or Nb with ferromagnets and SOC \cite{alidoust2015long,
alidoust2015spontaneous,
hikino2018magnetization, Bergeret2014,Jacobsen2015,Montiel2018,Eskilt2019,Bujnowski2019}.

 % Although possible in principle, the above mechanism of LRT generation requires specific type of SOC and superconductor/ferromagnet interface configuration.
 Without the supercurrent LRT are absent in the generic  S/F structures such as
shown in Fig.\ref{Fig:1}a.
Here the exchange field $\bm h\parallel \bm z$ produces only short-range triplets (SRT) with $S_z=0$
shown schematically by the blue arrows which decay at short  length of the order $\xi_F \sim 1$ nm in usual ferromagnets.  %
%%%%
%%%%
We demonstrate that generation of LRT with $S_z=\pm 1$
shown schematically by red arrows in Fig.\ref{Fig:1}b
can be achieved by inducing
 the superconducting condensate momentum
satisfying the condition $\bm h\times (\bm n\times \bm p_s)\neq 0$ .
This effect is fully controllable by the external magnetic field $\bm B \perp \bm h$ which induce the supercurrent  due to the Meissner effect $ \bm p_s   = - (\lambda_L/ \Phi_0)\bm n\times \bm B $ where $\lambda_L$ is the London penetration length and $\Phi_0$ is the magnetic flux quantum.

 The qualitative physics of the effect can be described as follows.
 We need three ingredients to generate LRT: spin-splitting field $\bm h$,
 SOC and supercurrent or condensate momentum $\bm p_s$.
The  role of $\bm h\parallel \bm z$ is to  generate $S_z=0$ spin triplet correlations.
The role of SOC is to convert them to
$S_z=\pm 1$ correlations  due to the momentum-dependent spin rotation.
 %%%
 The general structure of  anomalous function describing the pairing in spin-triplet channel can be parameterized\cite{leggett1975theoretical} as
$ (\hat{\bm\sigma}\cdot \bm d_{pw})$, where $\hat{\bm\sigma}$ is the vector of Pauli matrices. The $p$-wave spin vector is $\bm d_{pw} = F_{pw}(\omega) \bm h\times (\bm n\times {\bm p}) $
with the amplitude $F_{pw}(\omega)$ which is an even function of the Matsubara frequency $\omega$.
%and $\bm d_p = F_p(\omega) \bm h\times (\bm n\times {\bm p}) $

\begin{figure}[h!]
 \centerline{$
 \begin{array}{c}
 \includegraphics[width=1\linewidth]{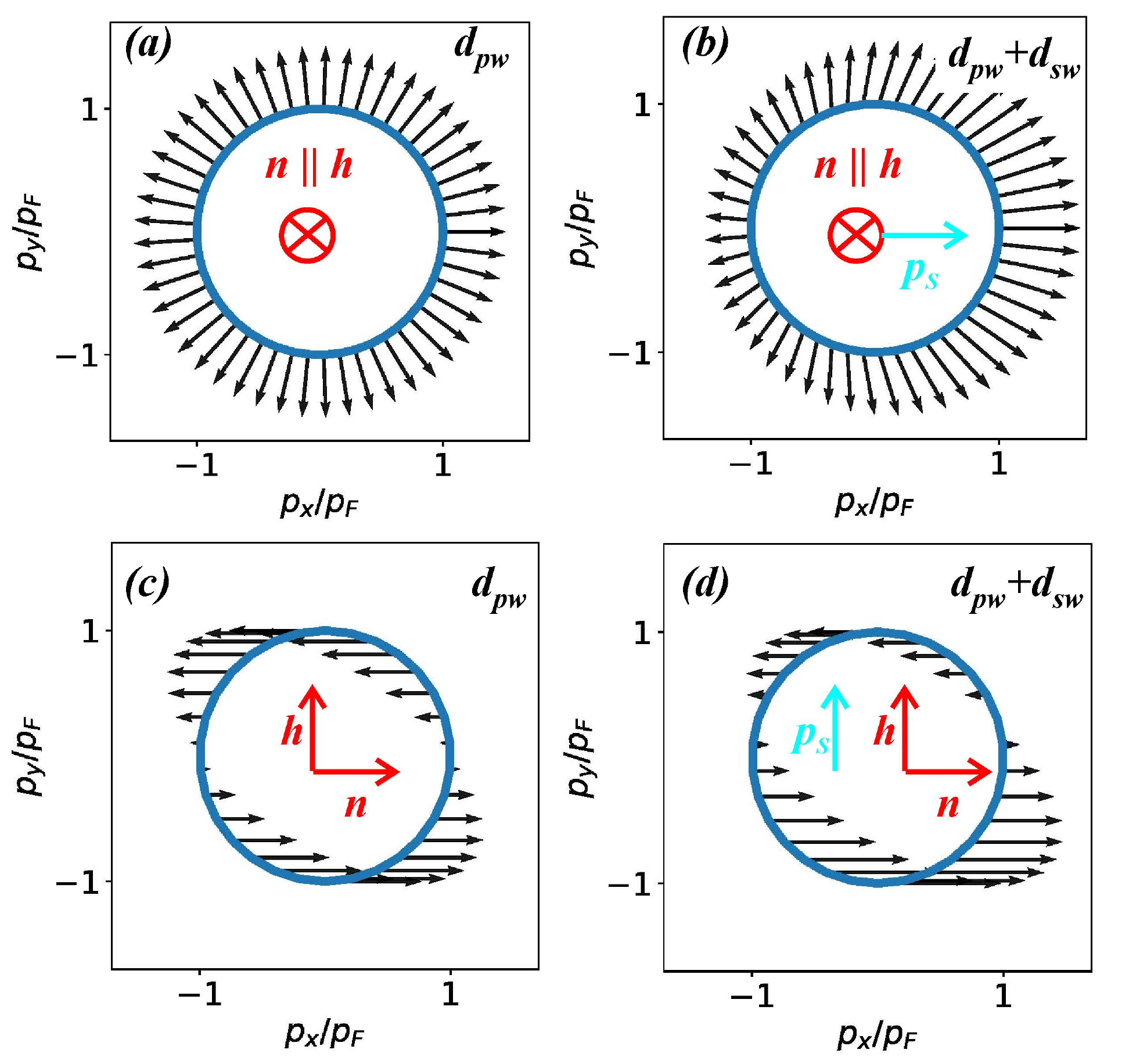}
 \end{array}$}
 \caption{\label{Fig:2}
   Textures of the spin-triplet order parameter vector $\bm d$
  on the Fermi surface $p_z=0$ cross-section. Left column:
  odd parity state $\bm d_{pw}$ for resting condensate $\bm p_s=0$. (a) $\bm n\parallel \bm h$,
  (c) $\bm n\perp \bm h$
  Right column: mixture of odd- and even-parity states $\bm d_{pw}+\bm d_{sw}$ induced by the moving condensate $\bm p_s\neq 0$. (b) $\bm h\parallel \bm n \perp \bm p_s$ and (d) $\bm h\parallel \bm p_s \perp \bm n$.   }
 \end{figure}

%%%
 By the order of magnitude
$ |\bm d_{pw}| \sim  h v_F\alpha /\Delta^2 \gg v_F\alpha /E_F$ which in principle is not  small. These triplets are even-frequency $\bm d_{pw}(\omega) = \bm d_{pw}(-\omega)$ and odd-parity p-wave $\bm d_{pw}(\bm p) = - \bm d_{pw}(-\bm p)$ correlations.
Depending on the relative orientation of the Zeeman field $\bm h$ and Rashba vector $\bm n $ the spin vector field $\bm d_{pw}$ can form different textures in momentum space. In Fig.~\ref{Fig:2}(a,c) the two characteristic examples of the hedgehog and domain wall are shown, respectively.

%The p-wave correlations can be observed in  ballistic SF systems \cite{chung2011topological,niu2012spin,PhysRevB.94.014515}.    However they are strongly suppressed in the presence of disorder and cannot  sustain long-range Josephson effect in diffusive systems. Here we show that in the presence of  externally induced superflow   $\bm p_s \neq 0$
% the spin-triplet p-wave correlations are partially transformed into the s-wave ones.

 %
 The  externally induced superflow $\bm p_s \neq 0$  induces Doppler shift of the quasiparticle energy levels\cite{PhysRevB.7.1001,kohen2006probing} $\bm v_F\cdot \bm p_s $.
 It results in the  suppression of pairing on one part of Fermi surface, namely for electrons with momentum $\bm p\parallel \bm p_s$.
In the simplest case of homogeneous system
this leads to the shift of imaginary frequencies so that the amplitude of triplet correlations is given by  $F_{pw}(\omega - i\bm v_F\cdot\bm p_s) \approx F_{pw}(\omega) - i(\bm v_F\cdot\bm p_s) \partial_\omega F_{pw}$.
This modification of the pairing amplitude
results in the additional component of the
spin vector $\delta \bm d = -i \partial_\omega F_{pw} (\bm v_F\cdot\bm p_s)\bm h\times (\bm n\times {\bm p})$.
Quite interesting for us is the s-wave component $\bm d_{sw} = \langle\delta \bm d \rangle_p$
given by $\bm d_{sw} = (2/3i)E_F (\partial_\omega F_{pw})  \bm h\times (\bm n\times {\bm p}_s)$.
The typical amplitude is
$|\bm d_{sw}|\sim  (p_s \xi) h v_F\alpha /\Delta^2 $ where $\xi$ is the coherence length.
%%%%
 Examples of spin textures  consisting of  the even and odd-parity mixtures $\bm d_{pw} + \bm d_{sw}$
 are shown in Figs.\ref{Fig:2}b,d.
 %%%
  %Since $\bm d_s \perp \bm h$ these correlations are equal-spin. In contrast to the p-wave component
  %the s-wave component is odd-frequency one.
  %
  To produce both the p-wave and s-wave triplets it is crucial to have $h\neq 0$ so that they are qualitatively different from those obtained beyond quasiclassical approximation due to the Edelstein effect \cite{Reeg2015,bobkova2017quasiclassical,PhysRevB.94.014515}.

 The described  mechanism has the same physical origin as the one reported in Ref.~\onlinecite{Eskilt2019},
where the gradients of pairing amplitude arising near the edges of S and Pt layers in the planar geometry
%the anomalous function  along the ferromagnet magnetization
in combinations with Rashba SOC can generate LRT.
  %%%
 In our proposal LRT are generated within the entire Pt layers thus providing much larger long-range Josephson currents. The other major advantage of our proposal is the full direct magnetic and electric control over the LRT generation because the condensate motion can be induced by applying external supercurrent or the external magnetic field.
 %Below we show that this leads to much larger Josephson currents mediated by LRT.
 %Below we quantify the described above mechanism and propose an experimental setup where it can manifested  through the long-range effect in the local density of states and a long-range Josephson effect switched by applying of the external magnetic field or a supercurrent parallel to the interfaces.

 %%%%%%%%%%%%%%%%%%%%%%%%%%%%%%%%%%%%%%
  We assume that SOC is small
  $v_F\alpha /E_F \ll 1$ which allows neglecting
  magnetoelectric effects \cite{Edelstein1990}
 and spontaneous phase gradients \cite{dimitrova2007theory,Buzdin2008,PhysRevB.92.125443}.
 Simultaneously this assumption allows using the quasiclassical theory describing the system in terms of the
  the quasiclassical   propagator
     $\check g = \check g (\omega,\bm p, \bm r)$, which is a $4 \times 4$ matrix in the direct product of particle-hole and spin spaces, $\omega$ is the imaginary frequency.
     It is determined by the Eilenberger equation  \cite{Eilenberger1968}
     %%%%%%%
  \begin{align} \label{Eq:Eilenberger}
  (\bm v_F \cdot  \hat{\bm\partial}) \hat g =
 [\check\Lambda, \hat g] + \tau_{imp}^{-1}
 [\langle \hat g \rangle_p , \hat g]
  \end{align}
 %%%
  Here $\bm v_F = \bm p/m$ is the Fermi velocity,
 $ \hat{\bm\partial}\hat g  =
 \bm \nabla \hat g  -
 i[\alpha \hat {\cal A} + \frac{\bm A}{2\Phi_0}  \hat\tau_3, \hat g]$ is the covariant gradient,
 $\hat \tau_i$ are Pauli matrices in particle-hole space.
 The quasiclassical equations are supplemented by the normalization
 condition $ \check g^2 =1$.
 We denote
 $\check \Lambda = \hat \tau_3 ( \omega   + i \bm h\cdot \hat{\bm \sigma} - i  \hat \Delta ) $, where $\hat \Delta =  |\Delta|  e^{i\hat\tau_3\chi} \hat\tau_1 $ is the superconducting order parameter, $\bm A$
 is the vector potential
 and  $\hat {\cal  A} $   is the general SU(2) field describing  SOC.
 The last term in Eq.~(\ref{Eq:Eilenberger}) describes impurity scattering with the rate $\tau_{imp}^{-1}$ and $\langle .. \rangle_p $ denote average over directions of $\bm p$.

 First, to elucidate the mechanism behind the generation of s-wave spin-triplets  let us consider solutions of Eq.~(\ref{Eq:Eilenberger}) with spatially homogeneous fields $|\Delta|$, $\bm h$, $\hat {\cal  A} $ and the gauge-invariant condensate momentum $\bm p_s=\bm \nabla \chi-\bm A/\Phi_0$.
 We consider expansion by small parameters  $v_F\alpha/\Delta \ll 1$ and $v_F p_s/\Delta \ll 1$.
  %%%%%%%%%%%%%%%%%
Let us denote as $\hat g_h$  the zeroth-order solution of Eq.~(\ref{Eq:Eilenberger}) with $\alpha=0$ and $\bm p_s=0$.
The first-order correction by SOC is \cite{supplement}
    \begin{align} \label{Eq:g1A}
 & \hat g_{\alpha} =
 i\alpha \hat g_h  \left[
 \hat g_h, \bigl(  \bm v_F\cdot \hat {\cal A} \bigr)
  \right]
 / (s_+ + s_- + 2 \tau_{imp}^{-1}),
 \end{align}
     %%%%%%%
   where
   $s_\pm = \sqrt{(\omega \pm ih)^2+\Delta^2}$.
This corrections generates pairing in  p-wave even-frequency spin-triplet channel
which is known to exist in superfluid $^3$He and systems with Rashba SOC\cite{Gor'kov2001}.
 For Rashba SOC $\hat {\cal A} = \bm n \times \hat{\bm\sigma}$
  the anomalous part of Eq.~(\ref{Eq:g1A}) is given by \cite{supplement}
  $  (\hat{\bm\sigma}\cdot \bm d_{pw})$ with the amplitude $F_{pw} = -i\alpha \Delta /(m s_+s_-(s_+ + s_- + 2 \tau_{imp}^{-1})) $.
      %%%
 By the order of magnitude
 $\hat g_{\alpha} \sim h v_F\alpha /\Delta^2$.
 %%%%%%%%%%%%%%%%%%%%%%%%%%%%%%%%%%%%%%%%%%%%%%%%%%%%

 As we discussed above the  condensate momentum adds Doppler shift to the Matsubara frequency in Eq.~(\ref{Eq:Eilenberger}).
   This results in the s-wave spin-triplet correction
  which in the clean limit is given by
  $\hat g_{\alpha A}= - (\bm v_F\cdot \bm p_s) \partial_\omega \check g_{\alpha}$ which contain the s-wave component
 with the spin structure described by the  vector $\bm d_{sw} \propto  \bm h\times (\bm n\times \bm p_s)$. By the order of magnitude
 $\hat g_{\alpha A}\propto ( \xi^2 \alpha h p_s/\Delta )  $
 where $\xi= v_F/\Delta$ is the coherence length in the ballistic limit $\tau_{imp}^{-1}=0$ and $\xi=\sqrt{D/\Delta}$ in the dirty limit $\Delta \tau_{imp}\ll 1$ where $D$ is the diffusion coefficient.

  The mechanism which we have discussed for the minimal model of  homogeneous system works as well for the proximity system where the spin splitting field is provided by the exchange interaction in the ferromagnet adjacent to the superconductor.
  The most striking demonstration of the  spin-triplet correlations with $\bm d_{sw}\perp \bm h$ can be achieved in a S/F structure with Rashba SOC shown in Fig.~\ref{Fig:1}.
   Below we show that generation of LRT in the  ferromagnet  can be fully controlled by the external magnetic field. This effect and its experimental manifestations are discussed below.

  %%%%%%%%%%%%%%%%%%%%%%%%%%%%%%%%%%%%%%
 {\it LRT Josephson effect in SFS junction stimulated by a magnetic field }
  For the setup in Fig.~\ref{Fig:1}a with  $\bm B=0$ we have $\bm d_{sw}=0$ and therefore despite the presence of SOC only   $S_z=0$ Cooper pairs are generated.  Such correlations decay in the diffusive ferromagnet at the length scale $\xi_F = \sqrt{D/h}$
which is rather short - $\xi_F\sim 1$ nm in usual materials.  Thus with exponential accuracy
the superconductivity is absent at the distances  $x\gg \xi_F$ from the S/F interface.

At the same time the correlations with $\bm d_{sw}\neq 0$
that appear at $\bm B\neq 0$ have $S_z=\pm 1$. Therefore they are LRT which are robust to the spin depairing and the only correlations that survive
 at the distances $x\gg \xi_F$ from the S/F interface in the ferromagnet. In the setup shown in Fig.\ref{Fig:1}a such pairs appear only upon applying the magnetic field $\bm B \perp \bm h$. Hence we claim to find the mechanism of the odd triplet superconductivity generated by the magnetic field. Formation of LRTs has important consequence in the transport properties\cite{
Robinson2010,Khaire2010,
singh2015colossal,Singh2016,
lahabi2017controlling,
Iovan2014}  which can
be directly measured using the electrical probes.
Below we discuss two of them - the tunnel conductance and Josephson current.

To describe the diffusive system we use Usadel equation for the s-component of the Green' function. It is obtained from the general Eilenberger equation (\ref{Eq:Eilenberger}) in the limit of large impurity scattering $\tau_{imp}h \ll 1$. Here we discuss  simplified version of the Usadel theory using following assumptions.
 First, we consider tunnelling S/Pt interfaces
 in Fig.\ref{Fig:1} so that superconducting correlations are small both in Pt and F. This allows to linearize the Usadel equation
 by assuming $\hat g = sign(\omega)\hat\tau_3 + \check f$ where the anomalous part $\check f= \hat f_s + \hat { \bm f}_t\cdot \hat{\bm \sigma}$ has spin-singlet $\hat f_s$ and spin-triplet $\hat {\bm f}_t$ components.

 Second, we assume that SOC is localized within Pt layers of the width
 $d_{soc}\ll \xi_N$, where $\xi_N = \sqrt{D/2\pi T}$.
  That is, integrating the Usadel equation through the Pt layer we get
 the boundary condition at the effective S/F interfaces \cite{supplement}
 %%%%%%%%
 \begin{align}
  n_x \nabla_x \hat f_s =  \gamma \hat F_{bcs}\label{bc_singlet}
  \\
  n_x \nabla_x \hat{\bm f}_t =  4 i  \tilde \alpha \hat\tau_3\hat{\bm f}_t\times (\bm p_s\times \bm n)\label{bc_triplet}
  \end{align}
  %%%%%%%%%
 where $\hat F_{bcs} = i\tau_3 \hat \Delta /\sqrt{\omega^2+\Delta^2}$ and the surface SOC strength $\tilde{\alpha} = \int dx \alpha (x)$ and $\gamma$ is the S/F interface transparency \cite{Kupriyanov1988, bergeret2005odd}.
 Note that the boundary condition  (\ref{bc_triplet}) corresponds to the conversion of p-wave to s-wave correlations shown in Fig.~\ref{Fig:2}.
 Indeed the r.h.s. of Eq.~(\ref{bc_triplet}) is equal to
  $  - i \int dx \langle
 (\bm p \cdot \bm p_s)  [\hat\tau_3,
 \hat g_{pw} ] \rangle_p
$,
where $\hat g_{pw}$ is the p-wave component of $\hat g$ and the integration is done over the Pt layer with $\alpha \neq 0$.

Finally, assuming that $p_s \xi\ll 1$ we neglect  the orbital depairing term $ p_s^2 D$ from the Usadel equation. That is only the linear in $\bm p_s$ term described by the boundary condition (\ref{bc_triplet}) is retained.
 Then, in the F layer where $\alpha =0$ and $\bm h\neq 0$ we obtain the linearized Usadel equaitons
  \begin{align}
     & \frac{D}{2} \nabla^2_x \hat f_s =
      |\omega|  \hat f_s  + i{\rm sgn}\omega (\bm h\cdot \hat {\bm f}_t) \label{usadel_singlet}
      \\
     & \frac{D}{2} \nabla^2_x \hat{\bm f}_t =
      |\omega|  \hat {\bm f}_t  +
     i{\rm sgn}\omega (\bm h \hat f_s) \label{usadel_triplet}
      %+ 4 i e\alpha \hat\tau_3\hat{ \bm f}_t\times (\bm A\times \bm n)
  \end{align}

 Solution of this system can be obtained in the compact form in the realistic limit
 $\xi_F\ll d_F\ll \xi_N$.  Further we are only interested in the LRT component of the anomalous Green's function which is approximately constant in the ferromagnet for the case under consideration.
  Up to the first order in the parameter $\tilde \alpha (p_s \xi_N)(\xi_N/d_F)$, which is assumed to be small, we obtain the LRT component in the ferromagnet\cite{supplement}
   \begin{align}
      \hat{\bm f}_t  =
       -\frac{\gamma\xi_F^2\tilde \alpha }{d_F|\omega| }
      \hat\tau_3 \hat F_{bcs}
      \bm h\times (\bm n\times \bm p_s)
      \label{ft_2}
  \end{align}
 Now we are ready to calculate the enhancement of  DOS in the ferromagnet due to the odd-frequency LRT
 \cite{Braude2007,PhysRevB.72.052512,DiBernardo2015}
 %%%%
  $ \delta N (\varepsilon) =
 \frac{1}{2}|\hat{\bm f}_t^2|(\omega = i\varepsilon + \delta ) $ where $\delta$ is the Dynes parameter.
    %%%%%
 For $d_F \gg \xi_F$ the contribution to this correction resulting from the short-range triplets already vanishes and, therefore, $\delta N = 0$ at $B=0$. However, at $B \neq 0$ the LRTs expressed by Eq.~(\ref{ft_2}) start to appear as $\delta N \propto B^2$.

%%%%%%%%%%%%%%%%%%%%%%%%%%%%%%%%%%%%%%%%%%%%%%%%%%%%
 \begin{figure}[h!]
 \centerline{$
 \begin{array}{c}
 \includegraphics[width=1\linewidth]{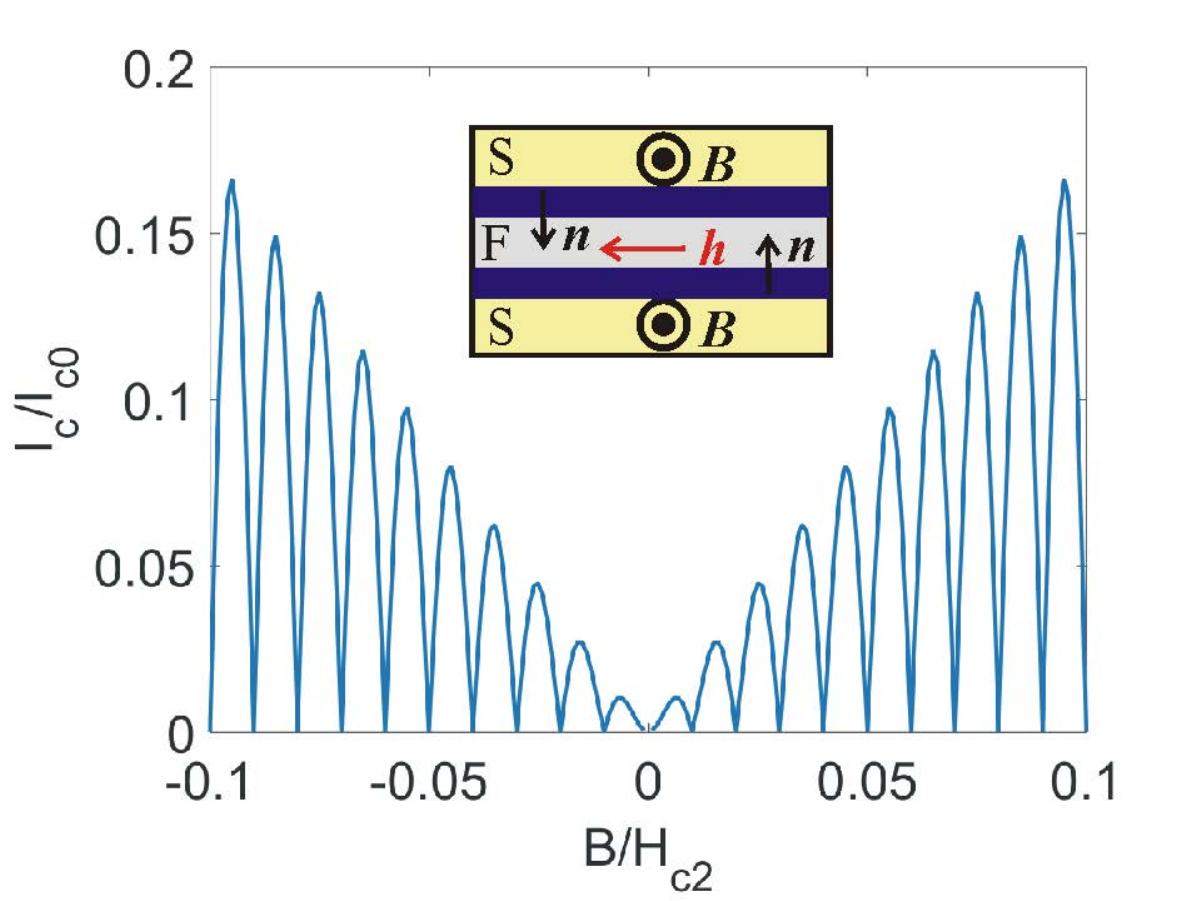}
 \end{array}$}
 \caption{\label{Fig:IcB}
  Interference patterns of the critical current $I_c(\Phi) $for the   magnetic-field induced Josephson effect through the magnetic and SOC interlayers as shown in inset. We use the parameters
  $T=0.1\Delta$, $d_F=\xi$, $L\lambda_L = 20 \xi^2$ and denote $H_{c2}=\Phi_0/\xi^2$.}
 \end{figure}
 %%%%%%%%%%%%%%%%%%%%%%%%%%%%%%%%%%%%%%%%%%%%%%%%%%%%
Next let us consider the long-range Josephson effect induced by magnetic field in the setup shown in the inset in Fig.\ref{Fig:IcB}. We assume similar conditions on the thickness of F layer $d_F$ as above.   %
 Having in hand the solutions (\ref{ft_2}) for
 left and right S/F boundaries
  we calculate \cite{supplement} the Josephson current-phase relation
 in the form $I= I_c\sin\phi$ with the critical current %
 \begin{align} \label{Eq:Ic}
  I_c /I_{c0} =
  \xi^2(\bm p_s \bm n_h)^2
  {\rm Re}\left[\psi \left(\frac{1+ia}{2}\right)-
    \psi\left(\frac{1}{2}\right) \right],
 \end{align}
   where $\bm n_h = \bm h/h$, $\psi$ is the di-gamma function,
  { $a=\Delta/\pi T$}, $ I_{c0} = 4\sigma_F \Delta (\gamma \tilde{\alpha} \xi_F)^2/e d_F$ and
      $\sigma_F$ is the ferromagnet conductivity.
  %$R = 1/(e\nu \gamma)$ is the surface resistance,
  %$\bm p_s$, $\bm n$ are taken either at right or left S/F interface
  %
 %Here we assume that
  %the product $|\bm p_s\times \bm n|$ is the same for the identical left and right  S/F interfaces.
  %
   The physical origin of unusual behaviour in Fig.\ref{Fig:IcB} is determined by the mechanism which is the main topic of the paper, namely generation of long-range spin-triplet correlations by the external magnetic field. As we have obtained in Eq.~(\ref{ft_2}), the amplitude of long-range spin-triplets is proportional to the condensate momentum
   $\bm p_s$  which is generated by the external magnetic field through the  Meissner effect.
   In the SFS junction with tunnel interface barriers and weak proximity effect \cite{galaktionov2003proximity}
    $ p_s (d_F/2)= -  p_s (-d_F/2) =  B \lambda_L/\Phi_0 $.
 Hence the amplitude of critical current grows as $I_c\propto B^2$ for small external magnetic field, when the total flux through the junction area $\Phi= 2\lambda_L L B$ is small $\Phi\ll \Phi_0$.  Here $L$ is the length of the junction.
  For larger fields  we need to take into account
 phase variation along the junction
 which leads to the usual factor $(L \sin\phi)/\phi$ in the critical current where $\phi= 2\pi \Phi/\Phi_0$.
 In result we obtain $I_c\propto B$ envelope dependence of the critical current shown in Fig.\ref{Fig:IcB}. This growth is bounded from above by the depairing effects
 from which the most important one is expected to be the vortex entry \cite{PhysRevLett.12.499} to the superconducting leads caused by the external magnetic field $B$. This happens at fields $H_v$ when the condensate momentum at the edge becomes of the order of the critical depairing value so that $p_s\xi \sim 1$.
 Depending on the level of disorder one can have the values \cite{PhysRevLett.12.499} $H_v/H_{c2} \sim 0.1 - 0.19$. For such samples the orbital effects can be neglected at fields $B<0.1 H_{c2}$ as in Fig.\ref{Fig:IcB}.

  The critical current magnitude
 can be estimated using Eq.\ref{Eq:Ic} with the typical junction parameters \cite{fominov2007josephson}.
 Assuming that the junction area is $50 \times 50 \;\mu{\rm m}^2$, $\sigma_F\sim (50\;  \mu\Omega\; {\rm cm} )^{-1}$,
 $d_F\sim \gamma^{-1} \sim 5 \xi_F$ and
 $D\sim 10\; {\rm cm^2/s}$,
 $h\sim 500\; {\rm K}$ so that
 $\xi_F \sim 3 {\rm nm}$.
 For the superconducting gap in Nb $\Delta \sim 10 {\rm K}$ so that the critical current is
 $I_c \sim 10^{-1}(p_s\xi)^2 \tilde \alpha^2$ A. Taking $\tilde \alpha \sim 0.1-1$
 \cite{lo2014spin,banerjee2018controlling, ast2007giant} and $p_s\xi=0.3$ we get the current $I_c \sim 10^{-2} - 10^{-4}$ A which is much larger than in SFS junctions where LRT are generated by the magnetic inhomogeneity\cite{fominov2007josephson}.
 %%%%%%%%%%%%%%
    This $I_c$ is much larger than obtained for the planar geometry \cite{Eskilt2019} because in our mechanism there are no physical limitations on the junction cross section.

The $I_c(B)$ pattern in Fig.\ref{Fig:IcB} drastically differs from the ones observed previously in non-ferromagnetic Josephson junctions  with SOC \cite{suominen2017anomalous,assouline2019spin} and ferromagnetic ones without SOC \cite{kemmler2010magnetic} .
This behaviour can be considered as the fingerprint of LRT produced due to the presence of interfacial Rashba SOC.
The direction of $\bm h$  in Fig.\ref{Fig:IcB}  can  be pinned tailoring the F layer shape anisotropy  so that the easy axis is perpendicular to $\bm B$.
Such scenarios were realized experimentally and coercive fields of about
 $0.3$ T
 %%%
for Fe samples were measured\cite{golikova2012double} which exceeds the field range in Fig.\ref{Fig:IcB} provided that $H_{c2} < 3$ T which is typical for Nb superconducting leads.

In conclusion, we have demonstrated a general mechanism which provides generation of odd-frequency spin-triplet superconductivity by moving the superconducting condensate in homogeneous ferromagnets. Physically the effect originates from a coupling of a Cooper pair spin with its orbital motion via SOC. The effect leads to appearance of LRT correlations in experimentally relevant S/F hybrids and long-range Josephson effect switchable by an externally applied magnetic field.

\begin{acknowledgments}
The work of M.A.S was supported by the Academy of Finland (Project No. 297439) and  Russian Science Foundation, Grant No. 19-19-00594.
The work of I.V.B and A.M.B has been carried out within the state task of ISSP RAS with the support by RFBR Grants No. 19-02-00466, 18-52-45011 and 18-02-00318. I.V.B. also acknowledges the financial support by Foundation for the Advancement of Theoretical Physics and Mathematics “BASIS”.
\end{acknowledgments}

\section*{Supplementary material for "Odd triplet superconductivity
 induced by the moving condensate"}
%DOS and spin-resolved DOS in spin-textured S/F bilayers

\subsection{Eilenberger equation and conversion between p-wave and s-wave spin-triplet correlations }

We use Eilenberger equation
 \begin{align} \label{AppEq:Eilenberger}
 & (\bm v_F \cdot \hat{\bm\partial}) \hat g =
 [\hat\Lambda, \hat g] + \tau_{imp}^{-1}
 [\langle \hat g \rangle_p , \hat g]
  \\
  \label{AppEq:GradGen}
 & \hat{\bm\partial} \hat g=
 \bm \nabla \hat g  -
 i[\alpha \hat {\cal A} + \frac{\bm A}{2\Phi_0}  \hat\tau_3, \hat g]
 \end{align}
 %%%

 %%
Here $\hat \tau_i$ and $\hat \sigma_i$ are Pauli matrices in particle-hole and spin spaces, respectively, $\tau_{imp}^{-1}$ is the impurity scattering rate. We denote
 $\check \Lambda = \hat \tau_3 ( \omega   + i \bm h\cdot \hat{\bm \sigma} - i  \hat \Delta ) $, where $\hat \Delta =  |\Delta|  e^{i\hat\tau_3\chi} \hat\tau_1 $ is the superconducting order parameter.

 In the presence of the phase gradient it is convenient to implement gauge transform
 $\hat g \to e^{i\hat\tau_3\chi/2}
 \hat g
 e^{-i\hat\tau_3\chi/2} $ which transforms the covariant gradient operator as follows
  \begin{align}
     \label{AppEq:GradGen1}
 & \hat{\bm\partial} \hat g=
 \bm \nabla \hat g  -
 i[\alpha \hat {\cal A} - \bm p_s \hat\tau_3, \hat g]
 \end{align}
 %%%
  where we introduce the gauge-invariant condensate momentum $\bm p_s=\bm \nabla \chi-\bm A/\Phi_0$.

 Below we consider analytical solutions of Eqs.~(\ref{AppEq:Eilenberger}),(\ref{AppEq:GradGen1}) in
 case of the homogeneous fields $|\Delta|$, $\bm p_s$
 and SOC $\alpha \hat {\cal A}$.
 We assume that condensate momentum $ p_s v_F/\Delta \ll 1$ and SOC $\alpha v_F/h \ll 1$ are small and use perturbation expansion by these parameters.

  %%%%%%%
  In the absence of condensate momentum and SOC
  the solution is given by
  \begin{align}
 & \hat g_h = \hat\sigma_0 \hat g_s
  + (\hat{\bm \sigma}\cdot  \bm h) \hat g_t
 \\
 & \hat g_{s} (\omega)= [\hat g_0(\omega + ih) +  \hat g_0(\omega - ih)]/2
 \\
 & \hat g_{t} (\omega)= [\hat g_0(\omega + ih) -  \hat g_0(\omega - ih)]/2h
  \end{align}
 the singlet and triplet components $\hat g_{s}$
 and $\hat g_{t}$ expressed through the
 GF in the absence of Zeeman field $ \hat g_0(\omega) = \hat \tau_3(\omega - i \hat \Delta )/s$
 and $s = \sqrt{\omega^2 + \Delta^2}$. This GF corresponds to the s-wave pairing and the Cooper pair spin lies in the plain perpendicular to the Zeeman field $\bm h$.

  %%%%%%%%%%%%%%%%%%%%%%%%%%%%%%%%%%%%%%%%%%%%
\subsection*{p-wave spin-triplet correlations due to  the Zeeman field and SOC}

 We search the correction $\hat g_{\alpha}$ of the first order in SOC from the following equation
 \begin{align} \label{AppEq:Eilenberger_2}
 & \begin{pmatrix}
  (s_+ + \tau_{imp}^{-1}) \hat g_+  & 0
  \\ \nonumber
   0 & (s_- + \tau_{imp}^{-1}) \hat g_-
 \end{pmatrix}
 \hat g_{\alpha}
 -
 \\
 & \hat g_{\alpha}
 \begin{pmatrix}
  (s_+ + \tau_{imp}^{-1}) \hat g_+  & 0
  \\
   0 & (s_- + \tau_{imp}^{-1}) \hat g_-
 \end{pmatrix}
 = i\alpha [ \hat g_h, (\bm v_F\cdot \hat {\cal A})]
 \end{align}
 We choose the spin quantization axis in $z$ direction and denote  $\hat g_\pm = \hat g_s \pm h \hat g_t$
 and $s_\pm = \sqrt{(\omega \pm ih)^2+\Delta^2}$.

 One can see that the solution satisfies anticommutation relation
 $\hat g_{h\alpha} (\bm h\cdot \hat{\bm \sigma}) +
  (\bm h\cdot \hat{\bm \sigma})\hat g_{h\alpha}  =0 $.
  Then we obtain the following solution
 %Taking into account that
 %%%
 %\begin{align}
 %\hat \Lambda^{-1} =
%\hat g_h  \frac{h(s_+ + s_-) - (\bm h\cdot \bm \sigma) (s_+ - s_-)}{2hs_+s_-}
% \end{align}
 %%%
%where $s_\pm = \sqrt{(\omega \pm ih)^2+\Delta^2}$, one can see that $\check g_{h\alpha}$ can be written in the form
\begin{align} \label{Eq:g1A}
 & \hat g_{\alpha} =
 i\alpha \hat g_h  \left[
 \hat g_h, \bigl(  \bm v_F\cdot \hat {\cal A} \bigr)
  \right]
 / (s_+ + s_- + 2 \tau_{imp}^{-1}).
 \end{align}

 { For Rashba SOC
 $\hat {\cal A} = \bm n \times \hat{\bm\sigma}$
Eq.~(\ref{Eq:g1A})  can be written in the form taking into account the explicit spin structure
 %%%%%%%%%%%%%
 \begin{align}
 \hat g_{\alpha} = \frac{\alpha
 ( \hat g_s\hat g_t + \hat g_t^2  \hat{\bm\sigma}\cdot\bm h )
 }{m h(s_+ + s_- + 2 \tau_{imp}^{-1})}
 (\hat{\bm\sigma} \cdot (\bm p \times \bm n)\times \bm h )
 \label{Eq:pwave}
 \end{align}
 %%%%%%%%%%%%%%%
 The contribution to anomalous part is given by the term ${\rm Tr}[\hat\tau_1\hat g_t\hat g_s] =-i \Delta /(2s_+s_-) $.  Hence the spin structure of the anomalous part can be written in the form $\hat{\bm\sigma}\cdot\bm d_p $ where the spin vector is $\bm d_p= F_p  \bm h\times (\bm p \times \bm n) $
 and the amplitude $F_p = i\alpha \Delta /(m s_+s_-(s_+ + s_- + 2 \tau_{imp}^{-1})) $. By the order of magnitude $|\bm d_p|\sim \alpha h v_F^2/\Delta^2$ in the clean limit and $|\bm d_p|\sim \alpha h v_F^2 \tau_{imp}/\Delta $ in the dirty limit. it can be written in the unified form as  $|\bm d_p|\sim \alpha h \xi^2$ where $\xi$ is the superconducting coherence length.
   }

\subsection*{Odd-frequency s-wave induced by the moving condensate }

  In general, the expressions for the correction induced by the condensate momentum $\bm p_s$ are rather lengthy. Simplified expressions can be obtained in the clean case $\tau_{imp}^{-1}=0$ and in the dirty limit what $\Delta\tau_{imp} \ll 1$ and
  $h\tau_{imp} \ll 1$.

   \subsubsection*{Clean limit}
  It the clean limit $\tau_{imp}^{-1}=0$
 we can obtain the solutions by adding the Doppler shift energy to the frequency $\omega - i\bm v_F\cdot \bm p_s$. The
 we can write the correction due to the SOC and condensate motion in the form
 \begin{align}
 \hat g_{ A\alpha} =
 -i (\bm v_F\cdot \bm p_s)
 \partial_\omega  \hat g_{ \alpha}
 \end{align}
 The s-wave component of the anomalous part is given by  $\hat{\bm\sigma}\cdot \bm d_s$ where the spin vector is $\bm d_{sw} = -i
 \langle (\bm v_F\cdot \bm p_s) \partial_\omega \bm d_{pw}\rangle_p $ or
\begin{align}
\bm d_{sw} =- (2i E_F/3) (\partial_\omega F_p)
\bm h\times (\bm p_s \times \bm n)
 \end{align}

\subsection*{Diffusive limit}

 The s-wave correlations are described by the momentum-average Green's function $\langle \hat g\rangle_p$. They are robust with respect to the impurity scattering so that they survive in dirty systems, which are more relevant to experiments.
 To describe them we use diffusive Usadel equation which is obtained from Eq.(\ref{AppEq:Eilenberger}) in the limit $\tau_{imp} \Delta \ll 1$
 \begin{equation}  \label{Eq:UsadelGen}  %\tag{S1}
 [\hat\Lambda, \hat g] = D\hat{\bm\partial}  ( \hat g  \hat{\bm\partial} \hat g)
 \end{equation}
 where $D= \tau_{imp} v_F^2 /3$ is the diffusion constant. Here we omit the angular brackets $\langle \hat g\rangle_p \to \hat g$.
  For a homogeneous superconductor in the presence of applied magnetic field or a supercurrent the action of this operator on the GF can be written in terms of the self-energy $[\check\Lambda + \check\Sigma, \check g] = 0$
  where
  the self-energy can be represented as a sum of three terms:
\begin{eqnarray}
\hat \Sigma = \hat \Sigma_{\alpha^2} + \hat \Sigma_{A^2} + \hat \Sigma_{A \alpha},
\end{eqnarray}
The components include
electromagnetic self-energy
$\hat \Sigma_{A^2} = (D/4) \bm p_s^2  \hat \tau_3 \hat g \hat \tau_3  $,
addition to spin relaxation
 $\hat \Sigma_{\alpha^2} = D \alpha^2 \hat {\cal A}_k \hat g \hat {\cal A}_k$,
 and  the most interesting for us part converting condensate motion into the deformation of the spin texture
 \begin{align}  \label{Eq:SelfEnergy}
 \hat \Sigma_{A\alpha} = -\frac{\alpha D}{2} \bm p_s
 (
 \hat{\cal {\bm A} } \hat g \hat\tau_3
  +
  \hat\tau_3
\hat g \hat{\cal {\bm A} }
 )
   \end{align}
%
%$\check \Sigma_{A\alpha} = D(\frac{e}{c}  A_k-\frac{\bm \nabla \chi}{2})  \hat \tau_3 \check g \hat {\cal A}_k + D\hat {\cal A}_k \check g (\frac{e}{c}  A_k-\frac{\bm \nabla \chi}{2})  \hat \tau_3$.

The correction to the GF due to the presence of the self-energy can be found also as a sum of the corresponding terms:
\begin{align} \label{Eq:gAExpansion_zero_field}
  \hat g =
  \hat g_h + \hat g_{\alpha^2} + \hat g_{A^2}+
  \hat g_{A \alpha}.
  \end{align}

 If we restrict ourselves by $\bm n \perp \bm n_h$, that is by the case on in-plane exchange field, then the corrections $\check g_{\alpha^2}$ and $\check g_{A^2}$ do not contain equal-spin correlations, which we are interested at the moment. Therefore, we only calculate $\check g_{A\alpha}$. Taking into account the normalization condition $\check g_h \check g_{A\alpha}+\check g_{ A\alpha}\check g_h = 0$ we obtain
the expression for this correction to GF:

 %%%%%%%%%%%%%%%%%%%%%%%%%%%%%

 { %%%%%%%%%%%%%%%%%%%%%%%%%%%
   \begin{align}
  \label{Eq:g1Aalpha}
 & \hat g_{A\alpha} =
 (\hat \Sigma_{A\alpha} - \hat g_h
 \hat \Sigma_{A\alpha} \hat g_h )/ (s_+ + s_-),
  \end{align}
 which can be rewritten in the form demonstrating the explicit spin structure of the GF:
 \begin{align*}  \label{Eq:swave_explicit}
  & \hat g_{A\alpha} = \frac{i\alpha D}{(s_++s_-)h} \times
  \\ \nonumber
 & ( \hat{\bm \sigma} \cdot \bm h\times (\bm p_s \times \bm n))
[ \Bigl(2\hat g_s^2 [\hat g_t,\hat\tau_3] +
 [\hat g_s,\hat g_t] \{\hat g_s,\hat \tau_3 \}\Bigr)
 +
 \\
 & (\hat{\bm \sigma}\cdot\bm h)
 \Bigl( 2\hat g_t^2 \{ \hat g_s,\hat\tau_3\} -
 [\hat g_s,\hat g_t] [\hat g_t,\hat \tau_3 ] \Bigr)
 ]
 \end{align*}

  Here the anomalous part is given by the first term in the brackets $\propto \hat\tau_1$
  and the normal part is the second term $\propto \hat\tau_0$.
 }

\subsubsection{Decay of p-wave correlations outside the Pt layer}

 Let is consider the linearized Eilenberger equation in the presence of impurity scattering and in the absence of SOC
 \begin{eqnarray} \label{Eq:Eilenberger}
 l_{imp} \partial_x\check g =
   [\langle \hat g\rangle, \hat g]
    \end{eqnarray}
 where angular brackets denote average over the directions of momentum and   $l_{imp}= \tau_{imp}v_F$ is the mean free path due to the impurity scattering. We assume that the
 impurity scattering rate is much larger than other frequencies.
  Assuming that superconducting correlations have p-wave symmetry we  can put
 $\langle \hat g\rangle  = sign(\omega) \hat\tau_3$.
Let us consider the $p_x$ component $ \hat f_x \cos\theta \hat\tau_1$.
The collision integral in Eilenberger Eq. yields admixture of the higher harmonics $\hat f_a \hat\tau_3\hat\tau_1 (\cos^2\theta - 1/2)$.
Using the fact that
$ \cos^3\theta = \cos\theta + \cos(3\theta)/2$ so that
$\cos\theta (\cos^2\theta - 1/2) = (\cos(3\theta) + \cos\theta )/2 $
we get the system of equations
 \begin{eqnarray} \label{Eq:Eilenberger}
 l_{imp} \partial_x\hat f_x =
   \hat f_a
   \\
    l_{imp} \partial_x\hat f_a =
  2 \hat f_x
    \end{eqnarray}
 which results in the second-order equation
 \begin{eqnarray}
      l_{imp}^2 \partial^2_x\hat f_x =
   2 f_x
 \end{eqnarray}
 Similar derivation can be done for $p_y$ and $p_z$ components. Hence one can see that p-wave
 correlations decay at the mean free path.
 
 \subsection{Derivation of the Usadel equation}
  In the limit of large impurity scattering we get from Eq.~(\ref{AppEq:Eilenberger}) the general relation between p-wave and s-wave components of the Green's function
 \begin{eqnarray} \label{AppEq:pWaveGen1}
   \hat g_{pw} (\bm p, \bm r) =
  \frac{\tau_{imp}}{2m}
  \hat g_{sw}
 (\bm p \cdot \hat{\bm\partial})
  \hat g_{sw}
 \end{eqnarray}
  %%%
  Here
 $\hat g_{pw}(\bm p, \bm r) = -\hat g_{pw}(-\bm p, \bm r) $ is the p-wave component of the Green's function,
  and $ \hat g_{sw}= \langle \hat g \rangle_p$ is the s-wave component.
  Substituting this relation back into the Eilenberger Eq.~(\ref{AppEq:Eilenberger})
  we get the general Usadel equation for s-wave component (here we omit the angular brackets)
  %%%%%%%%%%%%
 \begin{equation}  \label{Eq:UsadelGen}  %\tag{S1}
 D\hat{\bm\partial}  ( \check g
 \hat{\bm\partial} \check g) = [\check \Lambda, \check g]
 \end{equation}

 %%%%%%%%%%%%%%%%%%%%%%%%%%%
 \subsection*{Derivation of the boundary conditions}

In this subsection we derive the boundary condition (4) of the main text. Linearizing the general Usadel equation (\ref{Eq:UsadelGen}) with respect to the anomalous Green's function
and taking into account that in the considered Josephson geometry the anomalous Green's function depends only on the spatial coordinate $x$, we obtain the following equation for the anomalous Green's function:
\begin{eqnarray}  \label{Eq:Usadel_lin}  %\tag{S1}
 \frac{D}{2}\bigl\{ \partial_x^2 \check f +
 2 \alpha [\bm n \times \hat{\bm \sigma},
 \bm p_s \hat \tau_3 \check f]-
 i \alpha [\bm n \times \hat{\bm \sigma}, \partial_x \check f \bm x] -
 \nonumber \\
 \alpha^2 \bigl[ \bm n \times \hat{\bm \sigma},
 [\bm n \times \hat{\bm \sigma}, \check f] \bigr]\bigr\} =
 \nonumber \\
 \left(|\omega| + \frac{D p_s^2}{2}\right) \check f +
 \frac{i \; {\rm sgn}\;\omega}{2}
 \{ \bm h \hat{\bm \sigma}, \check f \}.~~~~~~~
  \end{eqnarray}
Further we integrate the above equation over the region $x \in (\mp d_F/2, \mp d_F/2 \pm d_{so})$ with $d_{so} \ll \xi_N$ near the  S/F interfaces where the SOC is nonzero. The results can be written as follows:
\begin{eqnarray}  \label{Eq:Usadel_lin_integrated}  %\tag{S1}
 \pm \frac{D}{2}
 \bigl( \partial_x \check f \bigl |_{\mp d_F/2 \pm d_{so}} - \partial_x \check f \bigl |_{\mp d_F/2} \bigr)
 = (|\omega| + \frac{D p_s^2}{2}) d_{so} \check f -
 \nonumber \\
 D \tilde \alpha [\bm n \times \hat{\bm \sigma},
 \bm p_s \hat \tau_3 \check f] +
 D \int \alpha^2 \bigl[ \check f -
 (\bm n \times \hat{\bm \sigma})
 \check f (\bm n \times \hat{\bm \sigma})\bigr]dx +
 \nonumber \\
 i \; {\rm sgn}\omega \; d_{so} [\hat {\bm f_t} \bm h +
 \hat f_s \bm h \hat{\bm \sigma}].~~~~~~~~~~~~~~
  \end{eqnarray}
Further we are interested in the triplet component of this equation and take into account that $\partial_x \hat {\bm f}_t \bigl |_{\mp d_F/2} = 0$. The r.h.s. of Eq.~(\ref{Eq:Usadel_lin_integrated}) contains only one term leading to the conversion of opposite-spin triplet correlations $\sim \bm h \hat{\bm \sigma}$ to the equal-spin triplet correlations with the $\bm d$-vector perpendicular to $\bm h$. It is the term $\sim [\bm n \times \hat{\bm \sigma}, \bm p_s \hat \tau_3 \check f]$. The other terms can probably be of the same order of amplitude, but they do not lead to the triplet-triplet conversion and, therefore, can be omitted because of their smallness with respect to the critical temperature. Neglecting these terms after the simple algebra we obtain Eq.~(4) of the main text.

%%%%%%%%%%%%%%%%%%%%%%%%%%%
\subsection{Solution of the linearized Usadel equations}

In this subsection we derive the expression for the LRT triplet component Eq.~(7) of the main text. The part of Eq.~(6) of the main text describing the LRT component with the $\bm d$-vector perpendicular to the exchange field takes the form:
\begin{align}
       & \frac{D}{2} \nabla^2_x \hat{\bm f}_{t,\perp} =
      |\omega| \hat {\bm f}_{t,\perp}   \label{usadel_triplet}
      %+ 4 i e\alpha \hat\tau_3\hat{ \bm f}_t\times (\bm A\times \bm n)
  \end{align}
  The characteristic length scale of $\hat {\bm f}_{t,\perp}$ is $\xi_N$. In the limit $\xi_F \ll d_F \ll \xi_N$ this equation can be integrated over $d_F$:
 \begin{eqnarray}
 \partial_x \hat {\bm f}_{t,\perp}\bigl |_{d_F/2} - \partial_x \hat {\bm f}_{t,\perp}\bigl |_{-d_F/2} = \frac{2|\omega|d}{D} \hat {\bm f}_{t,\perp}.
 \label{Eq:integrated}
 \end{eqnarray}
 Substituting the boundary condition (4) of the main text into the l.h.s. of Eq.~(\ref{Eq:integrated}) we obtain:
\begin{align}
 \frac{|\omega|d_F}{D} \hat {\bm f}_{t,\perp} = 2 \tilde \alpha i \hat \tau_3 \Bigl\{  [(\bm p_{s,R} \times \bm n_R) \times \hat {\bm f}_{t,R}]_\perp + \nonumber \\
 [(\bm p_{s,L} \times \bm n_L) \times \hat {\bm f}_{t,L}]_{\perp} \Bigr\}
\label{Eq:integrated_1}
\end{align}
Here subscripts $L,R$ denote if the corresponding quantities are taken at the left or right S/F interfaces.
In the framework of the perturbation theory up to the first with respect to the parameter $\tilde \alpha p_s \xi$ we should substitute into the r.h.s of Eq.~(\ref{Eq:integrated_1}) the value of the triplet anomalous Green's function at zero $p_s$. It is represented by the short-range triplet correlations and can be written as  $\hat {\bm f}_t = \hat f_{t,L(R)} \bm n_h$. The resulting expression for the LRT component is the following:
\begin{eqnarray}
 \hat {\bm f}_{t,\perp} = -\frac{2 \tilde \alpha  D i \hat \tau_3}{|\omega|d_F}(\hat f_{t,L} + \hat f_{t,R} )\bm n_h \times (\bm p_s \times \bm n)
\label{Eq:triplet}
\end{eqnarray}
The SRT anomalous Greens' functions $\hat f_{t,L(R)}$ decay at the short length scale $\xi_F$ from the S/F interface and, therefore, can be calculated at each of the interfaces separately according to the linearized Usadel equations Eqs.~(5-6) of the main text supplemented by the standard Kuprianov-Lukichev boundary conditions, which are expressed by Eqs.~(3-4) of the main text with $p_s=0$. Calculating $\hat f_{t,L(R)}$ according to this procedure  we obtain:
\begin{eqnarray}
 \hat {\bm f}_{t,L(R)}\bigl |_{\mp d_F/2} =  i\frac{\gamma \xi_F}{2} \hat F_{bcs} e^{\mp \phi/2}.
\label{Eq:triplet_short}
\end{eqnarray}
Substituting this expression into Eq.~(\ref{Eq:triplet}), for the S/Pt/F setup sketched in Fig.~1 of the main text  we obtain Eq.~(7) of the main text (due to the absence of the right S/F interface). For a Josephson junction geometry shown in the insert to Fig.~3 of the main text the both S/F interfaces generate SRT components and the resulting expression reads:
\begin{align}
      \hat{\bm f}_{t,\perp}  =
       -\frac{2\gamma\xi_F^2\tilde \alpha \cos \phi/2}{d_F|\omega| }
      \hat\tau_3 \hat F_{bcs}
      \bm h\times (\bm n\times \bm p_s)
      \label{ft_2_JJ}
\end{align}

 %%%%%%%%%%%%%%%%%%%%%%%%%
 \subsection{Details of the Josephson current calculation}

 For the problem under consideration the simplest way is to calculate the current at one of the S/F interfaces. Then it can be expressed as
 \begin{align}
 j=  i \gamma\frac{\sigma_F}{e} \pi T \sum_{\omega>0} {\rm Tr} \hat\tau_3(
 \hat {f}_s \hat F_{bcs})(x=d_F),
 \label{Josephson_interface}
\end{align}
where we need to calculate $\hat f_s$ up to the leading (second) order with respect to $\tilde \alpha (p_s \xi_N)$. The singlet and short-range triplet components decay rapidly into the depth of the ferromagnet and can be calculated at each of the S/F interfaces separately. For definiteness we consider the right S/F interface. The singlet and SRT anomalous Green's functions are to be calculated from Eqs.~(3)-(6) of the main text. In order to find them up to the second order with respect to $\tilde \alpha (p_s \xi)$ we need to substitute the first-order correction to the triplet anomalous Green's function Eq.~(7) of the main text into the right-hand side of Eq.~(6) of the main text with $\hat {\bm f}_{t}^{L,R} = (i\gamma \xi_F / 2) \hat F_{bcs}^{L,R}$. Then solving Eqs.~(3)-(6) we obtain the following result for $\hat f_s$:
\begin{eqnarray}
\hat f_s = -\frac{\gamma}{2}\xi_F \hat  F_{bcs}^R -
\tilde \alpha^2 (\bm p_s \bm n_h)^2\frac{4D\gamma |\Delta| \tau_2 \xi_F^2 \cos (\chi/2) }{d_F s |\omega|}.~~~~ \nonumber
\end{eqnarray}
Substitution of this result into Eq.~(\ref{Josephson_interface}) gives the sinusoidal current-phase relation with the critical current expressed by Eq.~(8) of the main text.

 %%%%%%%%%%%%%%%%%%%%%%%%%%%%%%%%%%%%%%%%%%%%%%
\subsection{Magnetic field screening and condensate momentum distribution in Josephson junction}

Let us consider two superconducting electrodes separated by the slit of the thickness $d_F$. Magnetic field in the slit, e.g. at $|x|<d_F/2$ is $B=B_0$
while in superconducting electrodes it is screened as
\begin{align}
  &  B= B_0 e^{(d_F/2-x)/\lambda_L} \;\; {\text for} \;\; x>d_F/2
    \\
 & B= B_0 e^{(x+d_F/2)/\lambda_L} \;\; {\text for} \;\; x<-d_F/2.
\end{align}

The vector potential satisfying $B= \partial_x A$ is given by

\begin{align}
 &   A = - \lambda_L B_0 e^{(d_F/2-x)/\lambda_L}  +  B_0(d_F/2+\lambda_L) \;\; {\text for} \;\; x>d_F/2
    \\
 & A =  \lambda_L B_0 e^{(x+ d_F/2)/\lambda_L} -  B_0(d_F/2+\lambda_L)   \;\; {\text for} \;\; x<-d_F/2
 \\
 & A = B_0 x \;\; {\text for} \;\; |x|<d_F/2
\end{align}
Because $A$ does not tend to $0$ at $x\to \pm \infty$ there are phase gradients $\nabla \varphi = \pm  B_0(d_F/2+\lambda_L)/\Phi_0$ which make the condensate momentum to vanish in the bulk. Hence the condensate momentum is
given by
\begin{align}
 &   p_s = - (\lambda_L B_0/\Phi_0) e^{(d_F/2-x)/\lambda_L} \;\; {\text for} \;\; x>d_F/2
    \\
 & p_s =  (\lambda_L B_0/\Phi_0) e^{(x+ d_F/2)/\lambda_L}  \;\; {\text for} \;\; x<-d_F/2
 \end{align}
 Near the interfaces $x= \pm d_F/2$ we have $p_s = \mp \lambda_L B_0/\Phi_0$.
 The phase gradients lead to the interference pattern in the magnetic field dependence of the critical current.

\bibliography{refs2}

\end{document}